\documentclass[pre,aps,pdflatex,
superscriptaddress,floats,showpacs,10pt]{revtex4-1}
\usepackage{graphicx}
\usepackage{amssymb,amsmath}
\usepackage{float}
\usepackage{xcolor}

\begin{document}

\title{On production and asymmetric focusing of flat electron beams using rectangular capillary discharge plasmas}

\author{G.\,A.\,Bagdasarov}
\affiliation{Keldysh Institute of Applied Mathematics RAS, Moscow, 125047, Russia}
\affiliation{National Research Nuclear University MEPhI (Moscow Engineering Physics Institute), Moscow, 115409, Russia}

\author{N.\,A.\,Bobrova}
\affiliation{Keldysh Institute of Applied Mathematics RAS, Moscow, 125047, Russia}

\author{A.\,S.\,Boldarev}
\affiliation{Keldysh Institute of Applied Mathematics RAS, Moscow, 125047, Russia}
\affiliation{National Research Nuclear University MEPhI (Moscow Engineering Physics Institute), Moscow, 115409, Russia}

\author{O.\,G.\,Olkhovskaya}
\affiliation{Keldysh Institute of Applied Mathematics RAS, Moscow, 125047, Russia}

\author{P.\,V.\,Sasorov}
\affiliation{Keldysh Institute of Applied Mathematics RAS, Moscow, 125047, Russia}

\author{V.\,A.\,Gasilov}
\affiliation{Keldysh Institute of Applied Mathematics RAS, Moscow, 125047, Russia}
\affiliation{National Research Nuclear University MEPhI (Moscow Engineering Physics Institute), Moscow, 115409, Russia}

\author{S.\,K.\,Barber}
\affiliation{Lawrence Berkeley National Laboratory, Berkeley, California 94720, USA}

\author{S.\,S.\,Bulanov}
\affiliation{Lawrence Berkeley National Laboratory, Berkeley, California 94720, USA}

\author{A.\,J.\,Gonsalves}
\affiliation{Lawrence Berkeley National Laboratory, Berkeley, California 94720, USA}

\author{C.\,B.\,Schroeder}
\affiliation{Lawrence Berkeley National Laboratory, Berkeley, California 94720, USA}

\author{J.\,van\,Tilborg}
\affiliation{Lawrence Berkeley National Laboratory, Berkeley, California 94720, USA}

\author{E.\,Esarey}
\affiliation{Lawrence Berkeley National Laboratory, Berkeley, California 94720, USA}

\author{W.\,P.\,Leemans}
\affiliation{Lawrence Berkeley National Laboratory, Berkeley, California 94720, USA}

\author{T.\,Levato}
\affiliation{Institute of Physics ASCR, v.v.i. (FZU), ELI-Beamlines Project, 182 21 Prague, Czech Republic}

\author{D.\,Margarone}
\affiliation{Institute of Physics ASCR, v.v.i. (FZU), ELI-Beamlines Project, 182 21 Prague, Czech Republic}

\author{G.\,Korn}
\affiliation{Institute of Physics ASCR, v.v.i. (FZU), ELI-Beamlines Project, 182 21 Prague, Czech Republic}

\author{M.\,Kando}
\affiliation{National Institutes for Quantum and Radiological Science and Technology (QST), Kansai Photon Science 
Institute, 8-1-7 Umemidai, Kizugawa, Kyoto 619-0215, Japan}

\author{S.\,V.\,Bulanov}
\affiliation{Institute of Physics ASCR, v.v.i. (FZU), ELI-Beamlines Project, 182 21 Prague, Czech Republic}
\affiliation{National Institutes for Quantum and Radiological Science and Technology (QST), Kansai Photon Science 
Institute, 8-1-7 Umemidai, Kizugawa, Kyoto 619-0215, Japan}
\affiliation{A.\,M.\,Prokhorov General Physics Institute RAS, Vavilov Str. 38, Moscow 119991, Russia}

\begin{abstract}
A method for the asymmetric focusing of electron bunches, based on the active plasma lensing technique is proposed.
This method takes advantage of the strong inhomogeneous magnetic field generated inside the capillary discharge plasma
to focus the ultrarelativistic electrons. The plasma and magnetic field parameters inside the capillary discharge are
described theoretically and modeled with dissipative magnetohydrodynamic computer simulations enabling analysis of the
capillaries of rectangle cross-sections. Large aspect ratio rectangular capillaries might be used to transport
electron beams with high emittance asymmetries, as well as assist in forming spatially flat electron bunches
for final focusing before the interaction point.
\end{abstract}


\maketitle

\section{Introduction\label{sec:intro}}

The laser acceleration of charged particles provides an approach towards the development of compact electron accelerators
with high energies and low emittance needed for coherent light sources and linear colliders~\cite{ECL-2009, LE-2009}.
Although there has been significant progress made recently in the understanding and production of the high energy and
quality electron beams via laser plasma interactions~\cite{ECL-2009, WPL-2014, Kim-2013}, the application of these beams
to coherent light sources and linear colliders is yet to be achieved. Linear electron-positron colliders need asymmetric
emittances and polarized electrons, with the smaller vertical emittance required to be of the order of 0.01\,mm mrad.
The asymmetric emittance, \textit{i.e.}, flat beams (see~\cite{Kando-2007} and references therein), are needed to
reduce the beam-induced synchrotron radiation (beam-strahlung~\cite{Chen-1992, Chen-Telnov-1989, Klasen-2002, Schroeder-2012})
at the interaction point. Since one of the laser plasma accelerator advantages is the short, compared to conventional ones,
acceleration length (the 4.25\,GeV electron beams were produced using a 9\,cm long capillary~\cite{WPL-2014}), it is
desirable important to keep the size of the transport and focusing sections of such accelerators of the same order.

Within the framework of the Laser Wake Field Accelerator concept~\cite{TD-1979}, M.\,Kando et al.~\cite{Kando-2007}
suggested a method for the electron injection via the transverse wake wave breaking~\cite{TWB-1997}, when the electron
trajectory self-intersection leads to the formation of an electron bunch elongated in the transverse direction thus
implying the flat electron beam generation.

Here we propose a plasma-based method for flat electron bunch formation, which is based on the active lensing
concept~\cite{Tilborg-2015, Tilborg-2017, Pompili-2017}, using a strong inhomogeneous magnetic field generated
in the capillary discharge plasma. The capillary discharges have applications in laser electron accelerators
as waveguides~\cite{Ehrlich-1996, Hosokai-2000, Bobrova-2001, ECL-2009, Gonsalves-2007, Kameshima-2009, WPL-2014,
Steinke-2016}, and for electron beam focussing~\cite{Tilborg-2015, Steinke-2016, Tilborg-2017}. The azimuthally
symmetric transverse magnetic field generated in the circular cross-section capillaries vanishes at the axis and
has a radially increasing strength. This  solenoidal field provides the electron beam focusing in both transverse
planes in contrast to the permanent-magnet quadruples. Moreover the magnetic field gradient of the permanent-magnet
quadruples approximately equals a few hundred of T/m (see Ref.\,\cite{Lim-2005}), whereas the capillary-discharge
field gradient can be ten times higher, reaching several kT/m~\cite{Pompili-2017}. We note that active plasma
lensing was demonstrated for ion beams in z-pinch plasma discharges~\cite{ION1, ION2, ION3,ION31} and in
non-pinching capillary discharges~\cite{ION4}.

Although the majority of the experiments uses circular cross-section capillaries, a square cross-section is more suitable
for several plasma diagnostic techniques~\cite{Gonsalves-2007, Jones-2003}. We also note that in contrast to the circular
cross-section capillaries, in a square cross-section capillary the magnetic field is not azimuthally
symmetric~\cite{Bagdasarov17}. It is plausible to assume that in the case of  rectangle cross-section
capillaries with large aspect ratio (which is the ratio of the long to short side length of the capillary in
the transverse plane), the magnetic field near the axis is almost one-dimensional. Its gradient along the short
capillary side is substantially larger than that along the long capillary side. It would result in different focus
lengths of the electrons (or ions) in different directions leading to the electron (ion) beam being first compressed
in the direction corresponding to the larger magnetic field gradient. We show in this work that the current distribution
inside the capillary supports this idea. One application of such magnetic system is the asymmetric focusing of charged
particle beams.

In this paper, rectangle cross-sections capillaries and charged particle focusing by the magnetic field inside such
capillaries are investigated. The paper is organized as follows. Section\,\ref{sec:MHDsetup} describes the configuration
and parameters of the magnetohydrodynamic (MHD) simulations. Section\,\ref{sec:MHDresults} contains the results of the
dissipative MHD simulations of the  rectangle, large-aspect-ratio, cross-section capillary. In Section\,\ref{sec:SIMPLEmodel}
we present analytical description of the plasma discharge density, temperature, and magnetic field distributions during the
quasi-stationary phase of the capillary discharge. The electron (ion) beam focusing by the capillary magnetic field
are considered in Section\,\ref{sec:ELfocus}. The scalings of the key capillary discharge parameters are given in
Section\,\ref{sec:SCALING}. Conclusions and discussions are presented in Section\,\ref{sec:CONCLUSIONS}.

\section{Simulation setup\label{sec:MHDsetup}}

To simulate the plasma dynamics and magnetic field evolution in the capillary discharge plasma we chose an oblong
rectangular capillary with the aspect ratio of 10:1 and the size of $[-L_x, +L_x]\times[-L_y, L_y]$ in the $(x,y)$-plane,
where $L_y = L = 125\,\mu$m and $L_x = 10\,L$. The capillary is pre-filled with pure hydrogen gas, which is initially
(at $t = 0$) distributed homogeneously, and its density is equal to $\rho_0 = 3.5\cdot10^{-6}\,\mathrm{g/cm^3}$,
corresponding to the electron density $n_e = 2.1\cdot 10^{18}\,\mathrm{cm^{-3}}$ when the gas is fully ionized.

\begin{figure}[h!t]
  \includegraphics[width=0.45\textwidth]{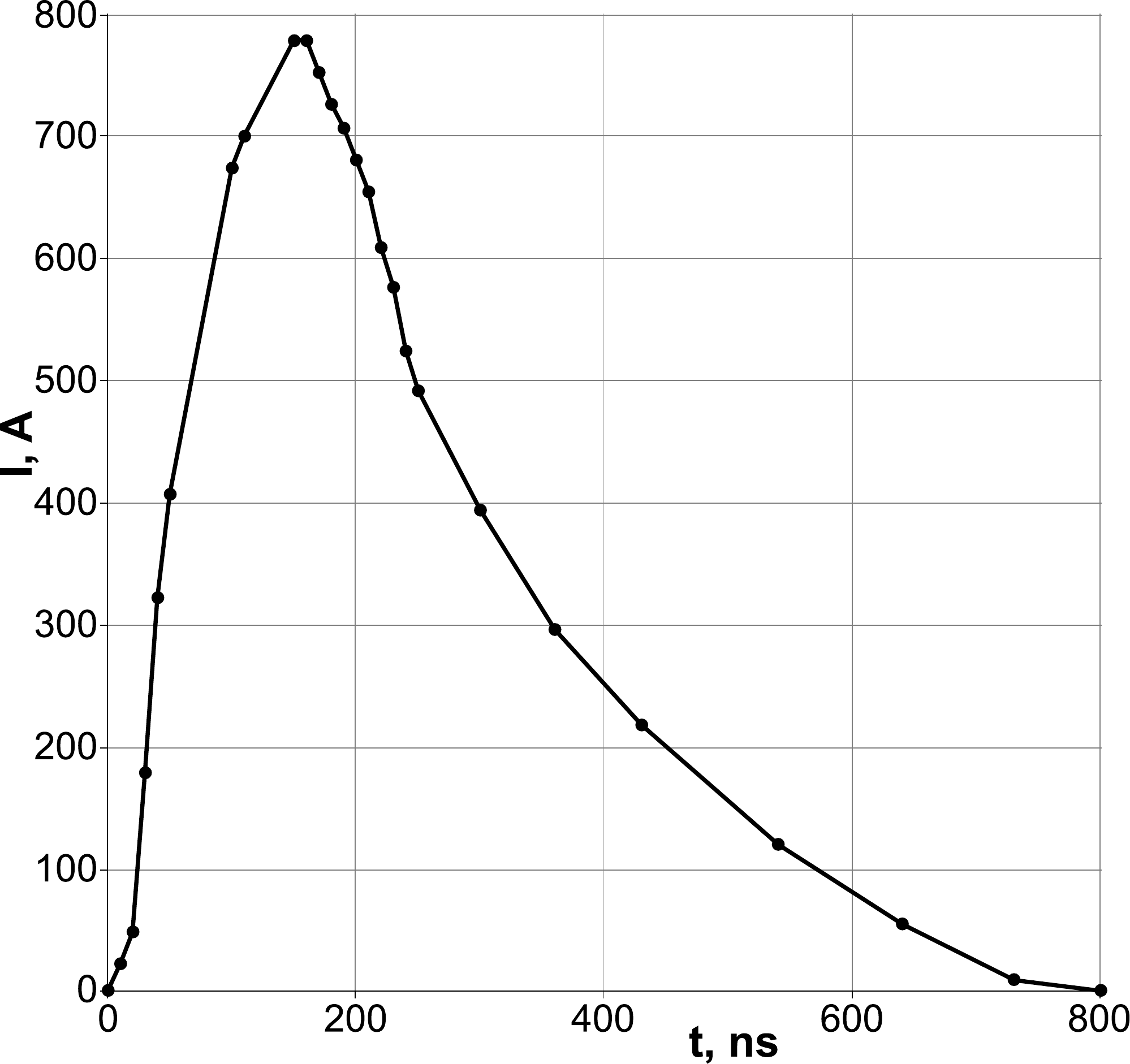}
  \caption{Time dependence of the electric current $I(t)$ inside the capillary.\label{Fig:It}}
\end{figure}

Initially there is no electric current inside the channel. The discharge is initiated by a pulse of current
 driven by an external electric circuit. The total electric current through the discharge, $I(t)$, is considered to be a 
given function of time. The current profile with the peak of 777\,A at 160\,ns is shown in Fig.\,\ref{Fig:It} and its form
is similar to experimental one in Ref.~\cite{WPL-2014}.. In 
order to initiate the discharge in the simulations, the
hydrogen is assumed to be slightly ionized ( $T_e = T_i = 0.5$\,eV) at $t=0$. 

For the sake of simulation simplicity, we calculate the evolution of the corresponding magnetic field implementing
an extended computational domain  with insulator around the capillary similarly to the case analyzed in
Ref.\,\cite{Bagdasarov17}. Total transverse size of the domain is $40\,L = 5$\,mm. The following MHD
equations are solved in this extended domain:

\begin{gather}
  \partial_t\rho + \nabla\left(\rho{\bf v}\right) = 0,\label{MHD1}\\
  \rho\left(\partial_t + {\bf v}\cdot\nabla\right){\bf v} = -\nabla p + \frac{1}{c}{\bf j}\times{\bf B},\label{MHD2}\\
  \nabla\times{\bf B} = \frac{4\pi}{c}{\bf j},\label{MHD3}\\
  {\bf j} = \sigma\left({\bf E} + \frac{1}{c}{\bf v}\times{\bf B}\right),\label{MHD7}\\
  \nabla\times{\bf E} = \frac{1}{c}\partial_t{\bf B},\label{MHD4}\\
  \nabla\cdot{\bf B} = 0,\label{MHD4a}\\
  \rho\left(\partial_t + {\bf v}\cdot\nabla\right)\varepsilon_e + p_e\nabla{\bf v} = \nabla\left(\kappa_e\nabla T_e\right) 
  + \frac{{\bf j}^2}{\sigma} - Q_{ie},\label{MHD6}\\
  \rho\left(\partial_t + {\bf v}\cdot\nabla\right)\varepsilon_i + p_i\nabla{\bf v} = Q_{ie},\label{MHD5}
\end{gather}
where $\rho$, $\mathbf{v}$, $p_e$ and $p_i$ are the plasma density, plasma velocity, and electron and ion pressures
respectively. In Eq.\,\eqref{MHD2} an artificial viscosity~$\alpha$ is substructed from the full
pressure~$p = p_e + p_i - \alpha$ in order to preserve monotony of the numerical solution. The electric and magnetic
fields, $\mathbf{E}$ and $\mathbf{B}$, obey Maxwell's equations~\eqref{MHD3}, \eqref{MHD4} and~\eqref{MHD4a} within
the framework of the quasi-static approximation (see Ref.\,\cite{LL-ECM}). In Eqs.~\eqref{MHD7} and~\eqref{MHD6},
$\sigma$ is the electric conductivity. 
The electron and ion specific energies are $\varepsilon_e$(includes the cost of ionization) and $\varepsilon_i$,
respectively. The electron thermal conductivity coefficient is equal to $\kappa_e$. The function $Q_{ie}$ describes the
energy exchange between the electron and ion components. The equation of state and dissipation coefficients take into
account the degree of the gas ionization similar to the method used in Ref.\,\cite{Bobrova-2001}. Here
$\kappa_{e}$, $\sigma$ and $Q_{ie}$ are given by the expressions
\begin{align}
  \kappa_e &= \frac{n_e T_e}{m_e\nu_{ei}}\,\Gamma_1(x_e,w),\label{eq:kappaheat}\\
  \sigma &= \frac{e^2 n_e}{m_e\nu_{ei}}\left[1 - \Gamma_5(x_e,w)\right]^{-1},\label{eq:sigma}\\
  Q_{ie} &= 3\frac{m_e}{m_i} n_e\nu_{ei}\left(T_e - T_i\right),\label{eq:Qie}
\end{align}
respectively. The functions $\Gamma_1(x_e,w)$ and $\Gamma_5(x_e,w)$ depend on the parameters $x_e = \omega_{Be}/\nu_{ei}$,
where $\omega_{Be} = eB/m_ec$ is the electron Larmor frequency, characterizing whether or not the electrons are magnetized,
and the Lorentz parameter $w$, showing the relative role of the electron-electron and electron-ion collisions
(for details see Ref.\,\cite{NAB-PVS-1993}). In the case under consideration the functions $\Gamma_1$ and $\Gamma_5$ are
of the order of unity. The electron-ion collision frequency $\nu_{ei}$ equals~to
\begin{equation}\label{eq:nuei}
  \nu_{ei} = \frac{4\sqrt{2\pi}e^4z^2 n_i\Lambda_{ei}}{3\sqrt{m_e}\,T_{e}^{3/2}},
\end{equation}
where $z$ is the mean ion charge and $\Lambda_{ei}$ is the Coulomb logarithm for electron-ion collisions,
which in the case under consideration is equal to
\begin{equation}\label{eq:CLei}
  \Lambda_{ei} = \frac{1}{2}\ln \frac{9 T_e^3}{4\pi z^2e^6 n_e(1 + z T_e/T_i)}.
\end{equation}
It is necessary to note, that at low temperatures, as long as the mean ion charge is considerably less
than unity, there is a noticeable fraction of neutral particles. Both, fully ionized plasma, and  partially ionized plasma, 
are treated  as a single two-temperature fluid. We assume that the conditions for the local thermodynamic equilibrium 
are satisfied separately for the electrons and ions. The equilibrium state of ionization, $z$, is determined by the
generalized to the case when $T_{i}\neq T_{e}$  Saha equation for a plasma composed of
atoms and singly-ionized ions. To take into account the partial ionization of hydrogen plasma
we re-normalize the electron-ion collision frequency $\nu _{ei}$ by
taking into account the contribution of neutral particles to the electron
scattering.  

The insulator medium is treated as immobile (${\bf v} = 0$) and with zero electric conductivity, $\sigma = 0$, hence
the electric current density vanishes, ${\bf j}=0$. 
Coupling between the external electric circuit and simulated discharge is determined 
by the boundary condition for the time-dependent azimuthal component of magnetic field 
at the boundary of simulated domain $r=\sqrt{x^2+y^2}=R_{ex}=20L$: $B_\varphi\left(R_{ex},t\right)=2 I(t)/(cR_{ex})$.
The simulation time is within the range, $t\in\left[0, 500\right]$\,ns. 
The MHD code MARPLE (Magnetically Accelerated Radiative PLasma Explorer)~\cite{MARPLE12, Bagdasarov3D, Bagdasarov17} is
used to perform the simulations.

\section{Simulation results\label{sec:MHDresults}}

\begin{figure}[h!t]
  \includegraphics[width=0.95\textwidth]{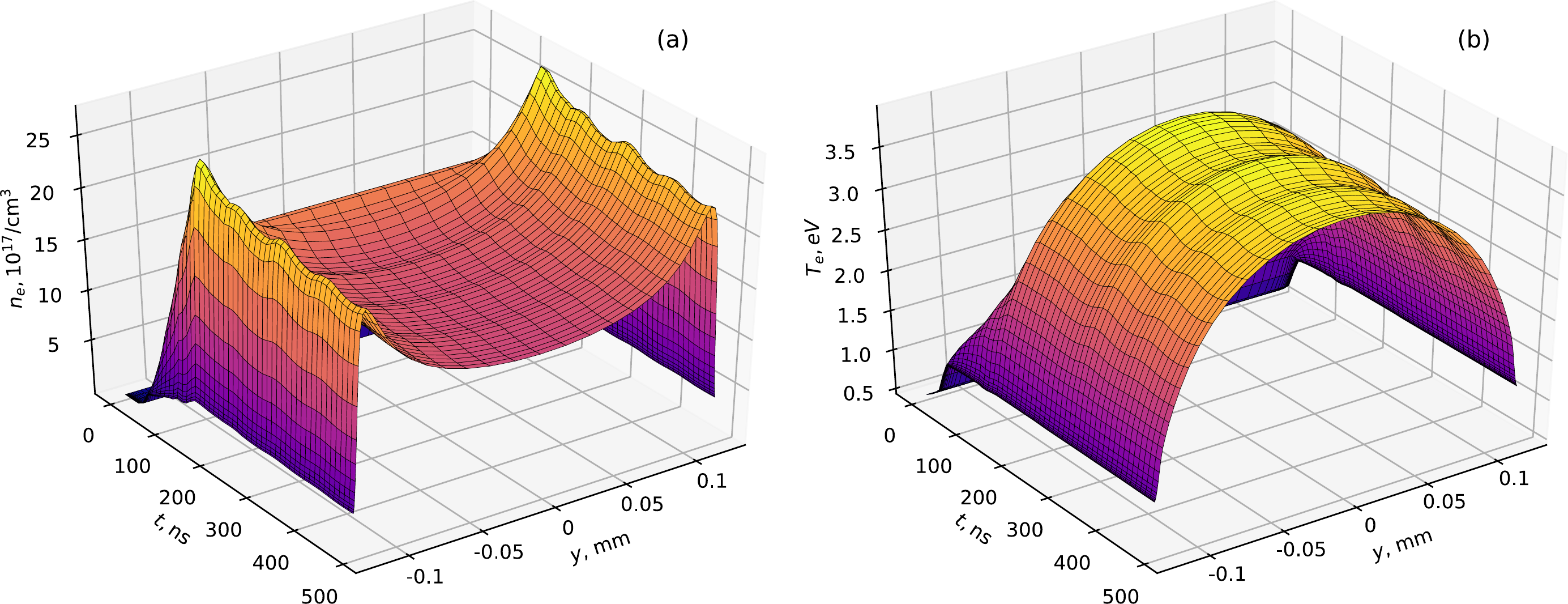}
  \caption{Time evolutions of electron (a) density and (b) temperature along $x = 0$ (solid black lines in
  Fig.\,\ref{fig:neTe}) for the range $y\in[-L, L]$.\label{fig:neTe-yt}}
\end{figure}

\begin{figure}[h!t]
  \includegraphics[width=0.95\textwidth]{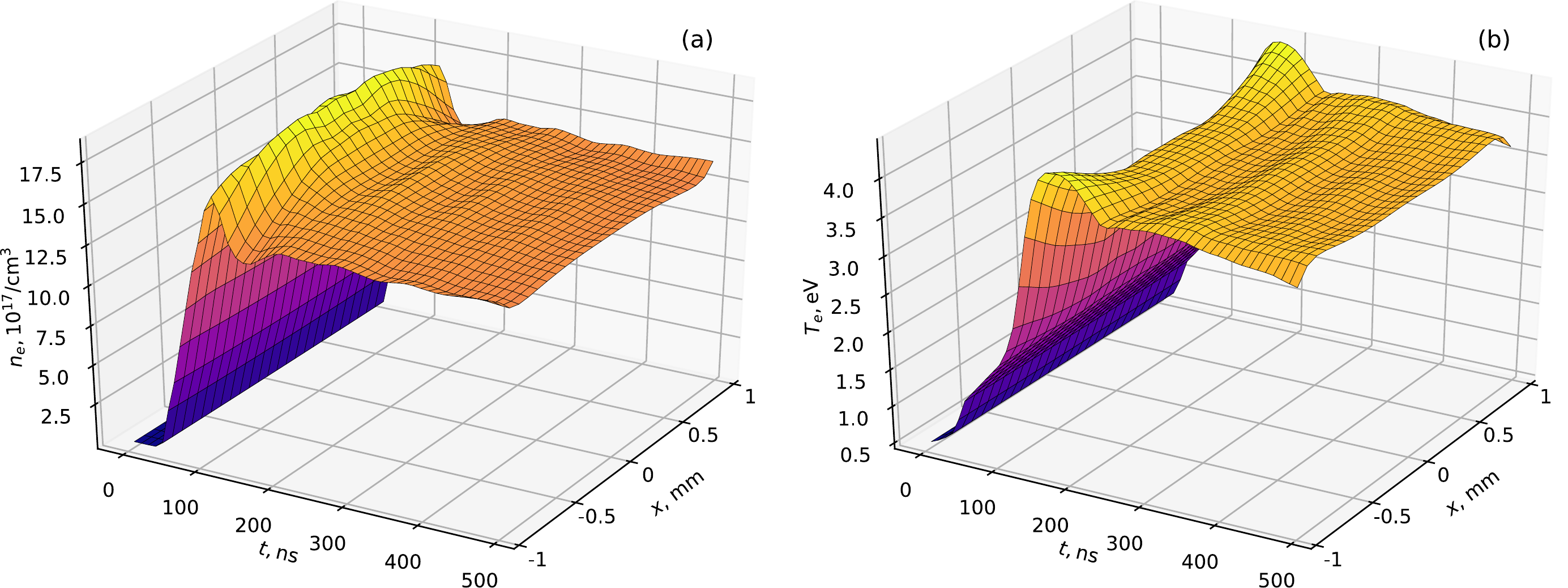}
  \caption{Time evolutions of electron (a) density and (b) temperature along $y = 0$ (dashed black lines in
  Fig.\,\ref{fig:neTe}) for the range $x\in[-1, +1]$\,mm.\label{fig:neTe-xt}}
\end{figure}

\begin{figure}[h!t]
  \includegraphics[width=0.95\textwidth]{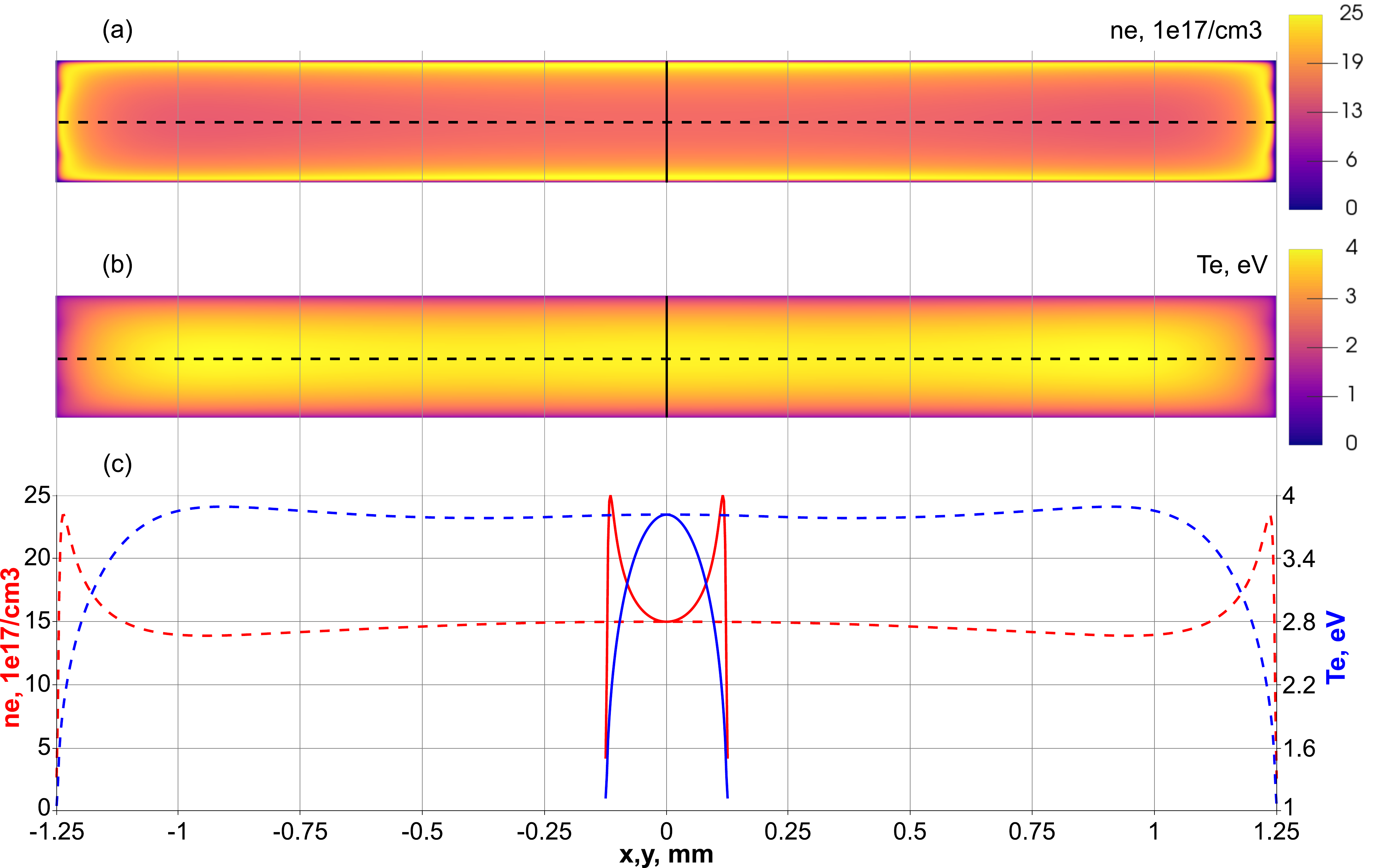}
  \caption{Distribution of the electron (a) density and (b) temperature at $t = 240$\,ns in the $(x,y)$-plane. (c)
  electron density distributions along the line $y = 0$ (dashed red line) and along the $x = 0$ (solid red line)
  and electron temperature vs $x$ along the $y = 0$ line (dashed blue line) and vs the coordinate $y$ along the
  $x = 0$ (solid blue line).\label{fig:neTe}}
\end{figure}

\begin{figure}[h!t]
  \includegraphics[width=0.95\textwidth]{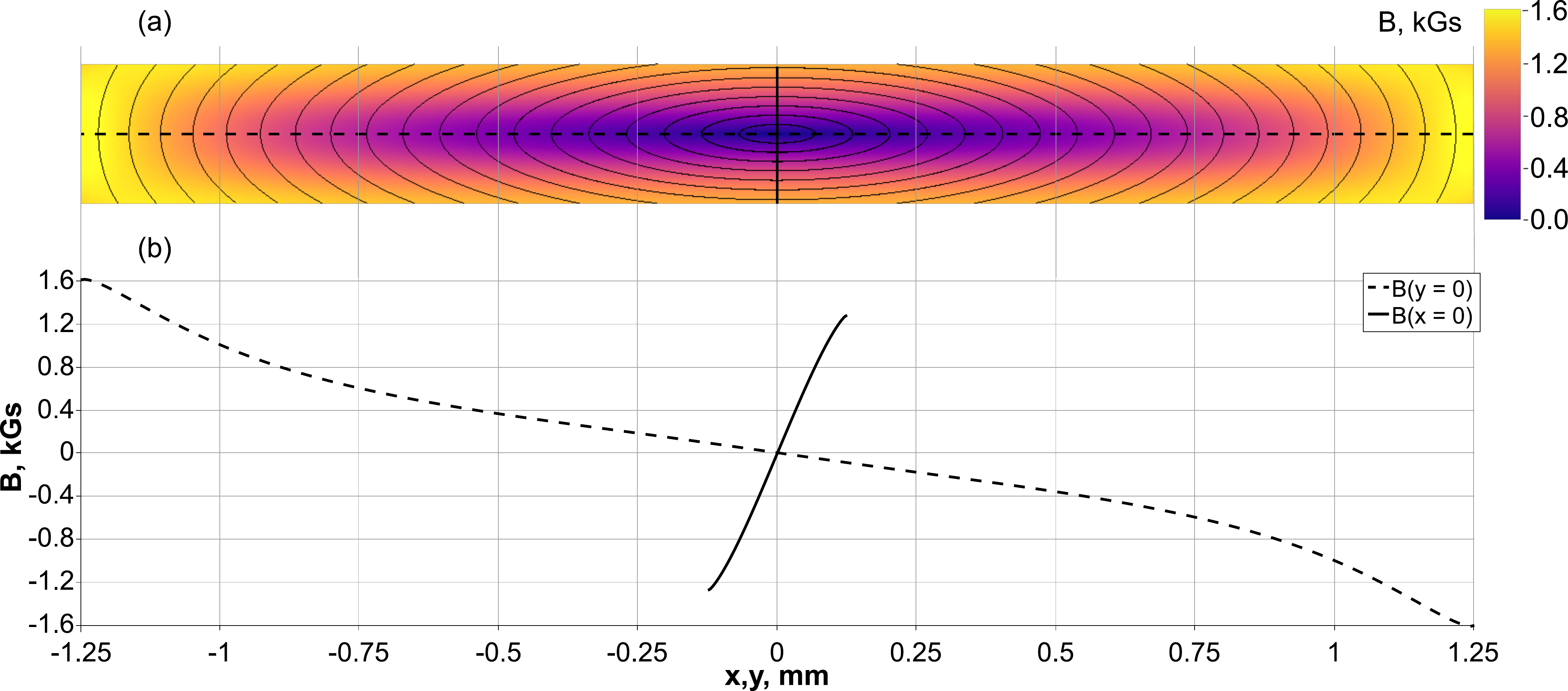}
  \caption{Distribution of the magnetic field at $t = 240$\,ns inside the capillary. (a) Magnetic field lines
  in the $(x,y)$-plane. (b) $B_x$ vs $y$ along the line $x = 0$ (solid black line) and $B_y$ vs $x$ along the
  line $y = 0$ (dashed black line).\label{fig:mfield}}
\end{figure}

The behavior of the discharge plasma in $y$-direction is quite similar to that of the circular and square capillary
plasmas~\cite{Bobrova-2001, Bagdasarov3D, Bagdasarov17} as can be seen from simulations. After a relatively
long period of plasma ionization ($\sim$\,100\,ns), plasma is heated by the Joule effect, however it remains
relatively cold in a small layer running along the capillary walls, which becomes more pronounced near the
shorter walls. This leads to formation of a plasma temperature maximum at the mid line of the slot and to the
redistribution of the plasma density in $y$-direction as shown in Fig.\,\ref{fig:neTe-yt}. A quasi steady state
of the discharge plasma is reached after approximately 100\,ns. At the steady state the plasma is at the mechanical
and thermal equilibrium, as well as magnetic field diffusion processes across the plasma are at the
equilibrium~\cite{Bobrova-2001}. This steady state evolves slowly due to the total electric current
changing in time and due to minor plasma dynamics in the $x$-direction. We see from Fig.\,\ref{fig:neTe-xt}
that a quasi-steady state is also reached after about 200\,ns in the $x$-direction. Parameters of such the
quasi steady state discharge are weakly dependent on the $x$ coordinate as can be seen in Fig.\,\ref{fig:neTe}.
At the quasi-equilibrium stage the magnetic field pattern corresponds to the asymmetric focusing of electron
bunches (see Fig.\,\ref{fig:mfield}).

In Fig.\,\ref{fig:neTe} we present the electron density and temperature distributions at $t = 240$\,ns after the
discharge was started. At this moment the variation of these parameters along the long side of the capillary
(i.e. along the dashed lines on the figure, $y = 0$) reaches its minimum.

We note that inside the capillary at some distance from its short sides, e.g. for $x\in[-1, +1]$\,mm,
the distributions of electron density and temperature along the line $y = 0$ are almost homogeneous.
The variation of the electron density $n_e$ is 7\% and the electron temperature $T_e$ variation is 3\% in
the defined region, i.e. the density and temperature depend mostly on the $y$-coordinate. As can be seen
from Fig.\,\ref{fig:neTe-xt}, the almost homogeneous distribution of the plasma parameters is reached after
approximately 200\,ns of the discharge and keeps constant during remaining 300\,ns of the simulation.
At the same time, as we see in Fig.\,\ref{fig:neTe-yt}, the quasi steady state along the $y$-coordinate
is reached after approximately 150\,ns and and is maintained till the end of the simulation.

We estimate the size of that region, where plasma parameters depend mostly on $y$-coordinate, as equal to
$[-8L, +8L]$. It equals to the length of the capillary long side $20L$ without two narrow regions of the
width approximately equal to the length of the capillary short side.

We assumed above that the magnetic field in the central part of an oblong capillary is almost one-dimensional,
i.e. its gradient along the short side is substantially larger than that along the long side. In Fig.\,\ref{fig:mfield}
we see that magnetic field dependence on the coordinates corresponds to this assumption.

The simulation gives the following dependence of the electron density $n_e$ and of the $B_x$ component of the magnetic
field on $y$ along the line $x = 0$ in the quasi-stationary regime:
\begin{align}
  n_e(0,y) &\approx n_e(0,0)\left[1 + 0.34\left(\frac{y}{L}\right)^2 + 0.20\left(\frac{y}{L}\right)^4\right],\label{3_010}\\
  B_x(0,y) &\approx y\cdot \partial_y B_x(0,0)\left[1 - 0.15\left(\frac{y}{L}\right)^2\right].\label{3_020}
\end{align}
Here the electron density at the capillary center equals $n_e(0,0) = 1.5\cdot 10^{18}\,\mathrm{cm^{-3}}$.
The magnetic field gradients along the $x$ and $y$ directions at $t = 240$\,ns, when $I = 524$\,A, are
\begin{align}
  \partial_x B_y(0,0) &= 0.73\,\mathrm{kGs/mm},\label{dBy}\\
  \partial_y B_x(0,0) &= 12.3\,\mathrm{kGs/mm}.\label{dBx}
\end{align}
As we see, the ratio $\partial_y B_x / \partial_x B_y \approx 16.85$ is of the order of the capillary aspect ratio,
which is equal to 10. We also note that accordig to the simulation results this ratio changes during the initial
stage of the discharge from zero to the quasi-stationary value, which may provide a tunable magnetic element for
achieving focusing and/or transport of electron beams with different degree of asymmetry. This property of the
rectangular capillary discharges will be investigated elsewhere.  The relationships~\eqref{3_010}, \eqref{3_020}
and~\eqref{dBx} obtained in simulations can be compared with Eqs.~\eqref{eq:nexi3}, \eqref{eq:Buh}
and~\eqref{eq:B_prime-pr}, respectively. The latter ones are obtained in the frame of a simple analytic theory
of the discharge presented in Sec.\,\ref{sec:SIMPLEmodel}. This comparison will be presented in
Sec.\,\ref{subsec:cap_params}.

\section{Plasma equilibrium and magnetic field profile\label{sec:SIMPLEmodel}}

The capillary discharge plasma under the conditions, when the magnetic Reynolds number $R_m = vL/\eta$,
where $v$ is the velocity scale of the plasma motion, $L$ is the space scale, and $\eta = c^2/4\pi\sigma$
is the magnetic diffusivity, is small, $R_m\ll 1$, the pinch effect is negligible, the axial electric field
is uniform across the capillary, the magnetic field pressure is much less than the plasma pressure, and hence
the plasma pressure can be considered to be constant across the capillary, the electrons are unmagnetized,
the electron and ion temperatures are equal to each other, $T_e = T_i = T$~\cite{Bobrova-2001}. In particular
these conditions are realized in the capillary schemes used  for the laser pulse guiding, when the plasma is
in the  dynamic equilibrium i.e. when the plasma velocity vanishes. As a result, the plasma density dependence
on the coordinate inside the capillary is determined by the balance between the Ohmic heating, $jE$ with $j$ and
$E$ being the electric current density and  the electric field, and the cooling due to the electron heat transport.
Here we formulate a theoretical model of the plasma equilibrium inside the flat capillary. The model in this case
is similar to that model, which has been formulated in Ref.\,\cite{Bobrova-2001} for the circular cross section capillary.

In the case of the capillary of large aspect ratio cross section, i.e. with one side considerably larger than the
other, we can assume that the parameters of the discharge at the equilibrium are almost independent of the $x$
coordinate, provided we exclude regions  of the size of the order of $L$ in the vicinities of the shorter sides
of the rectangle. Hence all parameters of plasma and magnetic field  depend  only on the $y$ coordinate. In this
case, the dependence of the plasma temperature on the $y$ coordinate is described by the heat transport equation
\begin{equation}
  \frac{d}{dy}\left(\kappa_e\frac{dT}{dy}\right) + \sigma E^2 = 0,\label{eq:heatT}
\end{equation}
where we use the relationship between the electric current density and the electric field, $j=\sigma E$.
the coordinate $y$.
Eq.\,\eqref{eq:heatT} is a nonlinear ordinary differential equation, which should be solved inside the
interval $y\in[-L, L]$ with the boundary conditions
\begin{equation}\label{eq:Tbound}
  \left. \frac{dT}{dy}\right|_{y=0} = 0\quad{\rm and}\quad\left. T \right|_{y=\pm L} = T_w.
\end{equation}
Here $T_w$ is the temperature of the capillary wall. For the sake of brevity we simplify the problem under
consideration as follows. Taking into account weak dependence of the Coulomb logarithm, $\Lambda_{ei}$, on
the temperature and density we assume that it is constant in Eqs.~\eqref{eq:kappaheat}, \eqref{eq:sigma}
and~\eqref{eq:nuei} with the typical value approximately equal to 3, retaining in the expressions for the
electron thermal conductivity \eqref{eq:kappaheat} and for the electric conductivity \eqref{eq:sigma} the
dependence on the temperature only. Using this assumption we rewrite Eqs.~\eqref{eq:kappaheat} and
\eqref{eq:sigma} in the form
\begin{equation}\label{eq:kappsigma}
  \kappa_e = \kappa_0 T^{5/2} \quad {\rm and}\quad \sigma = \sigma_0 T^{3/2}.
\end{equation}
Taking into account that the wall temperature is significantly lower than the plasma temperatute at
the capillary center we assume that $T_w = 0$.

Introducing the dimensionless variable
\begin{equation}\label{ex:ksi}
  \xi = y/L
\end{equation}
and the function
\begin{equation}\label{ex:uksi}
  u(\xi) = \left[U\,T(\xi)\right]^{7/2}
\end{equation}
with the parameter
\begin{equation}\label{ex:U}
  U = \sqrt{\frac{2\kappa_0}{7\sigma_0L^2E^2}}
\end{equation}
we obtain from Eq.\,\eqref{eq:heatT} the equation for the function $u(\xi)$
\begin{equation}\label{eq:uksieq}
  u''=-u^{3/7},
\end{equation}
which should be solved for $\xi\in[-1, 1]$ with the boundary conditions
\begin{equation}\label{eq:ubound}
  \left. u'\right|_{\xi=0} = 0 \quad {\rm and}\quad \left. u \right|_{\xi = \pm 1} = 0,
\end{equation}
which follow from Eq.\,\eqref{eq:Tbound}. Here and below a prime denotes differentiation with respect to the variable $\xi$.

Multiplying both sides of Eq.\,\eqref{eq:uksieq} by $u'$ and integrating over $\xi$ we obtain the integral
\begin{equation}\label{eq:uksint}
  \frac{u'^2}{2} = \frac{7}{10}\left(u_0^{10/7}-u^{10/7}\right),
\end{equation}
where $u_0$ is equal to the value of the function $u$ at the capillary center, $\xi=0$. The boundary
conditions~\eqref{eq:ubound} have been used. Using the integral~\eqref{eq:uksint} we can write the solution of
the boundary problem in quadratures:
\begin{equation}\label{eq:uquadr}
  \int\limits_0^{u/u_0}\frac{d s}{\sqrt{1 - s^{10/7}}} = \frac{1}{u_0^{2/7}}\sqrt{\frac{7}{5}}\left(1 - |\xi|\right).
\end{equation}
For the constant $u_0$ we have
\begin{equation}\label{eq:u0quadr}
  u_0 = \left[\sqrt{\frac{7}{5}}/\int\limits_0^{1}\frac{d s}{\sqrt{1 - s^{10/7}}}\right]^{7/2} = \left[\sqrt{\frac{7}{5\pi}}\frac{\Gamma(6/5)}{\Gamma(17/10)}\right]^{7/2} \approx 0.25,
\end{equation}
where $\Gamma(x)$ is the Euler gamma function~\cite{AS}.

The integral on the left hand side of the equation~\eqref{eq:uquadr} can be expressed via hypergeometric functions.
As a result the solution of the boundary problem~\eqref{eq:uksieq}-\eqref{eq:ubound} can be written in the implicit form
\begin{equation}\label{eq:uhyper}
  u(\xi)\,_2F_1\left[\frac{1}{2},\frac{7}{10},\frac{17}{10},\left(\frac{u(\xi)}{u_0}\right)^{10/7}\right] = u_0^{5/7}\sqrt{\frac{7}{5}}\left(1-|\xi|\right),
\end{equation}
where $u_0$ is given by Eq.\,\eqref{eq:u0quadr} and $_2F_1\left(a,b,c,x\right)$ is the Gaussian or ordinary
hypergeometric function~\cite{AS}.

\begin{figure}[h!t]
  \includegraphics[width=0.4\textwidth]{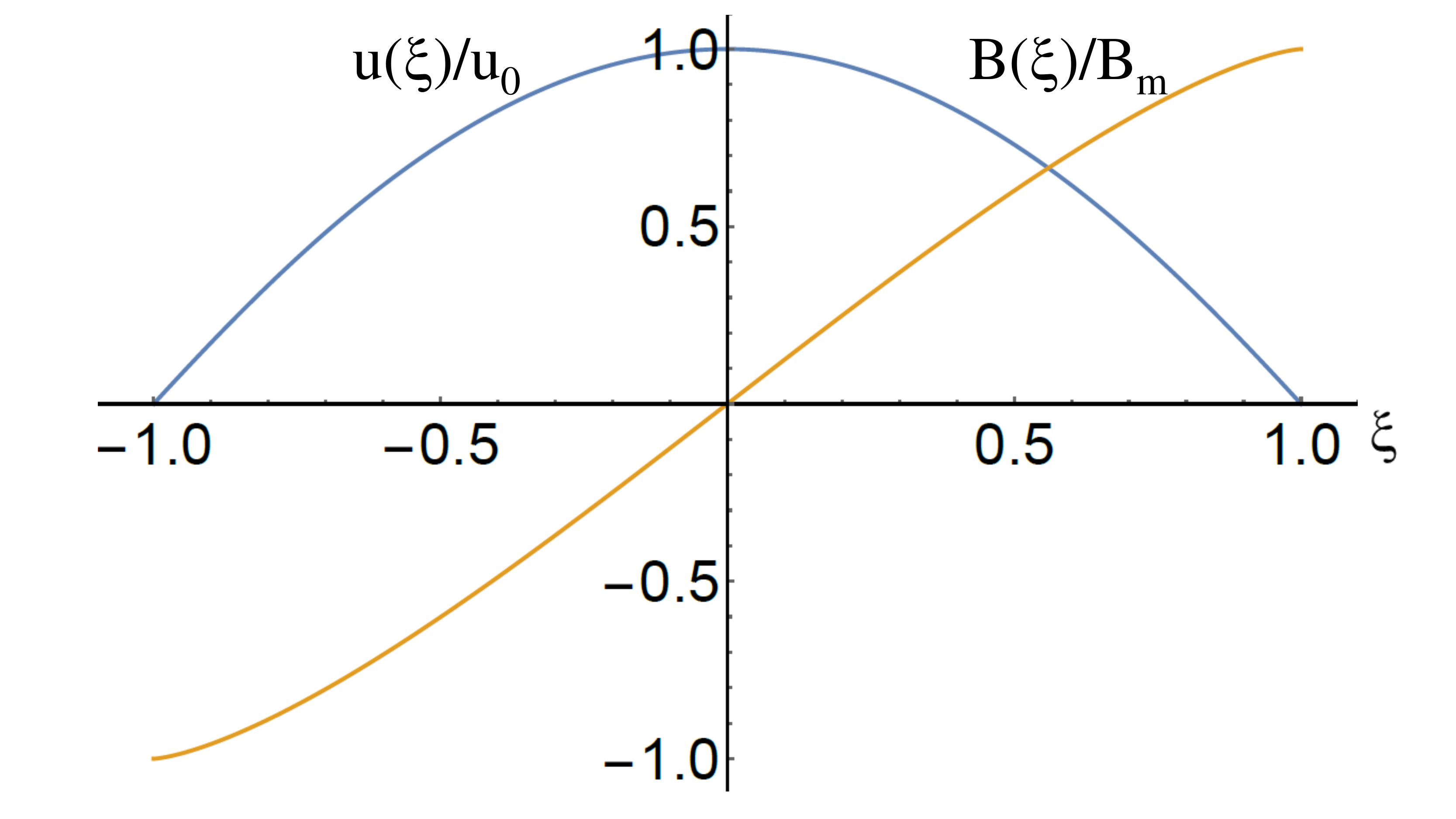}
  \caption{Dependence of the function $u(\xi)/u_0$ and of the normalized magnetic field $B(\xi)/B_m$ on the coordinate
  $\xi$ within the interval $[-1, 1]$.\label{fig:uBksi}} \end{figure}
Fig.\,\ref{fig:uBksi} shows the dependence of the function $u(\xi)/u_0$ on the coordinate $\xi$. Near the maximum the
function $u(\xi)$ can be approximated by the parabola. It is easy to find that for $\xi\ll 1$
\begin{equation}\label{eq:umax}
  u(\xi) = u_0\left(1 - \frac{1}{2 u_0^{4/7}}\,\xi^2 + \frac{1}{56\,u_0^{8/7}}\,\xi^4 + \cdots\right)
  \approx 0.25\left(1 - 1.10\,\xi^2 + 0.09\,\xi^4\right).
\end{equation}
Near the boundaries at $\xi\to\pm 1$  the function $u(\xi)$ depends on $\xi$ as
\begin{equation}\label{eq:umin}
  \frac{u(\xi)}{u_0} = \frac{1}{u_0^{2/7}}\sqrt{\frac{7}{5}}\left(1 - |\xi|\right) -
\frac{7}{34}\left[\frac{1}{u_0^{2/7}}\sqrt{\frac{7}{5}}\left(1 - |\xi|\right)\right]^{17/7} +\cdots
\end{equation}

The electron density distribution along the coordinate $\xi$ can be found from the condition of constant pressure
across the capillary. In the case of full ionization $z=1$ we have
\begin{equation}\label{eq:nexi1}
  p = 2n_e(\xi)T(\xi) = 2n_e(0)T(0)
\end{equation}
so that the electron density profile is given by
\begin{equation}\label{eq:nexi2}
  n_e(\xi) = n_e(0)\frac{T(0)}{T(\xi)} = n_e(0)\left(\frac{u_0}{u(\xi)}\right)^{2/7}.
\end{equation}
The electron density at the center is equal to
\begin{equation}\label{eq:nexi2a}
  \frac{n_e(0)}{\langle n_e\rangle} = \frac{10\,\Gamma(17/10)}{7\sqrt{\pi}\,\Gamma(6/5)}\approx 0.8,
\end{equation}
where
\begin{equation}\label{nemean}
  \langle n_e\rangle = (2L)^{-1}\,\int_{-L}^{L}n_e(y)\,dy
\end{equation}
is a mean electron density.

Near the axis $\xi=0$ it has the form
\begin{equation}\label{eq:nexi3}
  n_e(\xi) = n_e(0)\left(1 + \frac{1}{7 u_0^{4/7}}\cdot\xi^2 + \frac{2}{49\,u_0^{8/7}}\cdot\xi^4 + \cdots\right) \approx
  n_e(0)\left(1 + 0.31\,\xi^2 + 0.20\,\xi^4\right).
\end{equation}
The magnetic field, ${\bf B} = B(\xi){\bf e}_x$, where ${\bf e}_x$ is the unit vector in the $x$ direction,
should be found from the equation~\eqref{MHD3}, which takes the form
\begin{equation}\label{eq:Bxi1}
  B' = \frac{4\pi\sigma_0 T^{3/2} E L}{c}.
\end{equation}
Using the relationship~\eqref{ex:uksi}, we represent Eq.\,\eqref{eq:Bxi1} as
\begin{equation}\label{eq:Bxi1a}
  B' = h_1\left(\frac{u}{u_0}\right)^{3/7}.
\end{equation}
The normalized gradient of the magnetic field near the capillary axis is given by
\begin{equation}\label{eq:h0}
  h_1 = \frac{4\pi\sigma_0 E L u_0^{3/7}}{c\,U^{3/2}} =
\frac{7^{3/2} 2^{5/4}\,\pi^{1/4}}{5^{3/4}}\,
\left[\frac{\Gamma(6/5)}{\Gamma(17/10)}\right]^{3/2}\,
\frac{\sigma_0^{7/4} L^{5/2} E^{5/2}}{c\,\kappa_0^{3/4}}
\approx 17.82\,\frac{\sigma_0^{7/4} L^{5/2} E^{5/2}}{c\,\kappa_0^{3/4}}.
\end{equation}
It follows from Eqs.~\eqref{eq:uksint} and~\eqref{eq:Bxi1a} that the magnetic field dependence on the coordinate
$\xi$ has the following form:
\begin{equation}\label{eq:Bu}
  B(\xi) = \sqrt{\frac{7}{5}}\,u_0^{2/7}\,h_1\,\sqrt{1 - \left(\frac{u(\xi)}{u_0}\right)^{10/7}}.
\end{equation}
This relationship can be written as (we note that $u^\prime$ is negative at $\xi>0$)
\begin{equation}\label{eq:Buprime}
  B(\xi) = -\frac{h_1}{u_0^{3/7}} u'(\xi).
\end{equation}

Fig.\,\ref{fig:uBksi} presents the dependence of the normalized magnetic field $B(\xi)/B_m$ on the coordinate $\xi$.
As we see, the magnetic field vanishes at the axis, $\xi = 0$, where $u(\xi)|_{\xi=0} = u_0$.
Expanding expression~\eqref{eq:Bu} near $u = u_0$, i.e. for $\xi\ll 1$, we obtain the following
expression for the magnetic field
\begin{equation}\label{eq:Buh}
  B = h_1\,\xi - h_3\,\xi^3 + \dots \approx B^\prime(0)\,\xi \left(1 - 0.16\,\xi^2\right)
\end{equation}
with $h_3 = h_1/14 u_0^{4/7}$. The magnetic field maximum is reached at the capillary boundary,
$\xi = 1$, where $u(\xi)|_{\xi=1}=0$, and is equal to
\begin{equation}\label{eq:Bmax}
  B_m = h_1\sqrt{\frac{7}{5}}\, u_0^{2/7}.
\end{equation}
The total (per unit width) electric current in the capillary, $I_0 = c B_m/2\pi$, is
\begin{equation}\label{eq:Itot}
  I_0 = \frac{c h_1}{2\pi}\left(\frac{7}{5}\right)^{1/2}u_0^{2/7} =
  \frac{7^{5/2} 2^{1/4}}{5^{7/4}\, \pi^{5/4}}\,\left[\frac{\Gamma(6/5)}{\Gamma(17/10)}\right]^{5/2}\,\frac{\sigma_0^{7/4} L^{5/2} E^{5/2}}{\kappa_0^{3/4}} \approx
  0.32\,\frac{\sigma_0^{7/4} L^{5/2} E^{5/2}}{\kappa_0^{3/4}}.
\end{equation}
The electric current can also be derived from Eq.\,\eqref{eq:uksieq}:
\begin{equation}\label{eq:Itotuprime}
  I_0 = \int\limits_{-L}^Lj\, dy=\sigma_0 E L\int\limits_{-1}^1T(\xi)^{3/2}\, d\xi =
\frac{\sigma_0 E L}{U^{3/2}}\int\limits_{-1}^1u(\xi)^{3/7}d\xi=-\frac{2\sigma_0 E L}{U^{3/2}}u'|_{\xi=1}.
\end{equation}
It follows from Eq.\,\eqref{eq:uksint} that $u'|_{\xi=1} = -\sqrt{7/5} u_0^{5/7}$. Substituting $U$ and $u_0$ given by
the Eqs.~\eqref{ex:U} and~\eqref{eq:u0quadr}, respectively, into the right hand side of Eq.\,\eqref{eq:Itotuprime} we
obtain the expression \eqref{eq:Itot} showing that the electric current $I_0$ is proportional to the electric field in
the power of $5/2$ as in the case of circular capillary discharges~\cite{Bobrova-2001}.

\section{Charged particle focusing\label{sec:ELfocus}}

In what follows we consider the motion of an electron in the magnetic field of the rectangular capillary discharge
in the framework of a simplified one-dimensional model. We neglect any wakefield effects and assume electrons to be
probe particles. The trajectory of a charged particle moving in a constant one-dimensional magnetic field can be
found from the conservation of the particle energy and of the $z$-component of the generalized momentum,
\begin{equation}\label{eq:gamma-Pz}
  \sqrt{1 + p_x^2 + p_y^2 + p_z^2} = \gamma_0 \quad {\rm and} \quad p_z+A_z(\xi) = p_{0,z}.
\end{equation}
Here $\gamma = \sqrt{1 + p_x^2 + p_y^2 + p_z^2}$ is the relativistic Lorentz factor and $A_z$ is the $z$-component
of the vector potential ${\bf A} = A_z {\bf e}_z$. The $x$ component of the momentum, $p_x = p_{0,x}$, is constant.
These integrals of the particle motion follow from the independence of the magnetic field on time and $z$-coordinate.
The normalizations of the momentum on $m_e c$ and of the electromagnetic potential on $m_e c^2/e$ are used.
The magnetic field is equal to the curl of the vector potential, ${\bf B} = \nabla\times{\bf A}$, i.e. $B = A'_z(m_e c^2/eL)$.

Eqs.~\eqref{eq:Bu} and~\eqref{eq:uksint} yield the relationship between the $z$-component of the electromagnetic
potential and the function~$u(\xi)$:
\begin{equation}\label{eq:Azu}
  A_z(\xi) = A_0\,u(\xi)
\end{equation}
with
\begin{equation}\label{eq:A0}
  A_0 = -\frac{e h_1 L}{m_e c^2 u_0^{3/7}}.
\end{equation}
Using equations for the particle coordinates, $\dot \xi = p_y/\gamma$ and $\dot \zeta = p_z/\gamma$ with $\xi$ given by
Eq.\,\eqref{ex:ksi} and $\zeta = z/L$, where a dot stands for the time derivative, and Eq.\,\eqref{eq:uksint}, we can
obtain the dependence of the particle normalized $\zeta$-coordinate on the $\xi$-coordinate in quadrature:
\begin{equation}\label{eq:zuxi}
  \zeta(\xi) = \zeta_0 + \frac{1}{u_0^{5/7}}\sqrt{\frac{5}{7}}\,\int\limits_{u_0}^{u(\xi)}\frac{(p_{0,z} - A_0u)\,du}
  {\sqrt{\left[1-(u/u_0)^{10/7}\right]\left[\gamma_0^2-1-p_{0,x}^2-\left(p_{0,z}-A_0u \right)^2 \right]}}.
\end{equation}

Since in the region near the z-axis for $\xi\ll 1$ the magnetic field can be approximated by a linear function of
the $\xi$-coordinate the aberration in the focusing of high energy electrons can be made weak. This implies a
smallness of two parameters,
\begin{equation}\label{eq:del1del2}
  \delta_1 = \frac{eh_1a_0^2}{p_{0,z} m_e c^2 L} \quad {\rm and} \quad
  \delta_3 = \frac{h_3a_0^2}{h_1 L^2},
\end{equation}
where $a_0$ is approximately equal to the initial coordinate $y_0$. The first small parameter  shows that in
the limit $\delta_1\ll 1$ the electron with large enough momentum along the $z$-axis, $p_{0,z}$ or/and small
initial position $y_0$ moves inside the nonadiabatic region, where it is not magnetized. The electron oscillates
in the $y$-direction around the position $y = 0$ and moves with approximately constant velocity $v_z\approx c$ along
the $z$-axis. In addition, the smallness of the parameter $\delta_1$ indicates the role of the oscillation anharmonism
due to the nonlinear dependence of the $z$-component of the electron momentum on the coordinate even in the case
of the magnetic field linearly dependent on $y$. The second parameter $\delta_3$ should be small to make insignificant
the effects of the nonlinear term in the magnetic field dependence on the coordinate~\eqref{eq:Buh}.

Small but finite anharmonizm results in the dependence of the oscillation frequency on the amplitude, which leads to
the intersection of the particle trajectory with the $y=0$ axis at different coordinates $z$. This leads to the
magnetic lens aberration. The electrons being focused do not meet after the lens in one focal point. The further
the trajectories initially are from the axis, the closer to the lens they intersect the axis. It is convenient to
illustrate this by considering the electron motion near the axis in the magnetic field given by Eq.\,\eqref{eq:Buh}.
Using smallness of the electron displacement from the axis and retaining the terms of the order not higher than $y^3$,
the equations of electron motion can be written in the form of the anharmonic oscillator equation
\begin{equation}\label{eq:elmot}
  \frac{d^2 y}{dt^2} + \omega_0^2 y = \kappa y^3
\end{equation}
with the oscillation frequency and the parameter of nonlinearity equal to
\begin{equation}
  \omega_0 = \sqrt{\frac{e h_1 p_{0,z}}{m_e L \gamma_0^2}} \quad{\rm and}\quad
  \kappa = \frac{e}{m_e L^3 \gamma_0^2}\left(\frac{e h_1^2 L}{2 m_e c^2} + h_3 p_{0,z}\right),
\end{equation}
respectively. The nonlinearity parameter can also be written as $\kappa = (\omega_0^2/a_0^2)(\delta_1/2 +
\delta_3)$ with $\delta_1$ and $\delta_3$ given by Eq.\,\eqref{eq:del1del2}.

We consider the electron focusing by two magnetic configurations, by a ``long focusing system'' and by a ``short focusing
system''. In the first case, the particles are focused at or near the capillary end. In the case of the short focusing
system, the focus is located far from the capillary end.

\begin{figure*}[h!t]
  \includegraphics[width=0.95\textwidth]{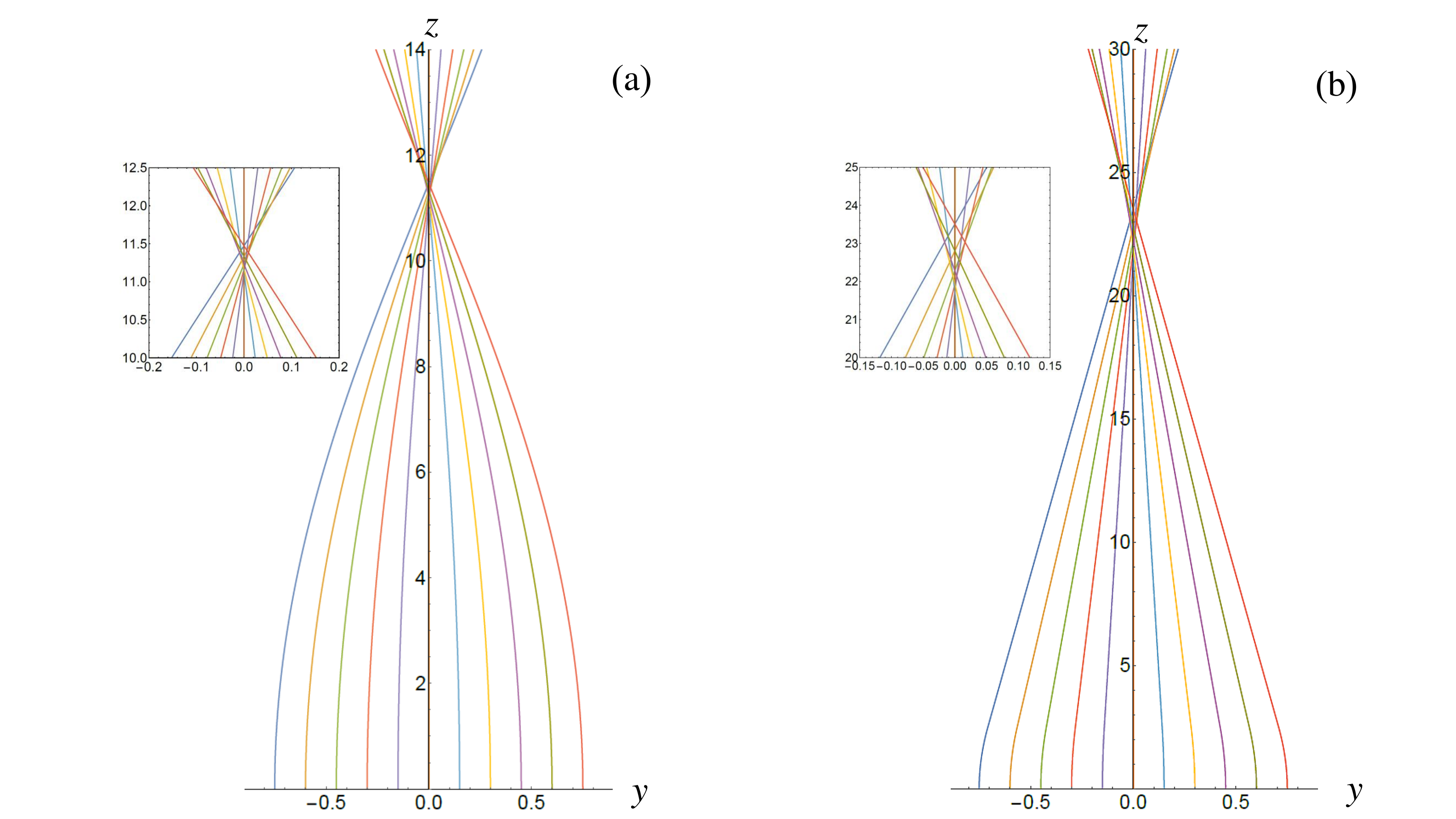}
  \caption{Electron trajectories in the magnetic field~\eqref{eq:Buh}. (a) Long magnetic focusing system. The inset
  shows the close-up of the focus region. (b) Short magnetic focusing system. The inset shows the close-up of the
  focus region.\label{FIG:focus}}
\end{figure*}

In the long focusing system the particle trajectory $y|_{t=z/c}$ is given by the solution of Eq.\,\eqref{eq:elmot},
which to the third order of the small oscillation amplitude is~\cite{LL-M}
\begin{equation}\label{eq:yt}
  y(t) = a_0 \cos{\left(\omega_0t-\frac{3 \kappa }{8 \omega_0}a_0^2 t\right)} +\frac{a_0^3\kappa }{32 \omega_0^2}\cos{3\omega_0 t}.
\end{equation}
It yields the relationship between the amplitude $a_0$ and the initial coordinate $y_0$, which reads
\begin{equation}
  a_0\approx y_0-\frac{y_0}{32}\left(\frac{\delta_1}{2}+\delta_3 \right).
\end{equation}
According to Eq.\,\eqref{eq:yt}, the time, at which the trajectory intersects the axis $y=0$, is approximately equal to
\begin{equation}
  t \approx \frac{\pi}{2\omega_0} + \frac{3 \pi \kappa }{16 \omega_0^2}y_0^2.
\end{equation}
It depends on the initial coordinate $y_0$, which leads to the finite focus width  equal to
\begin{equation}
  \Delta z = \frac{3\pi\kappa c}{16\omega_0^2}Y_0^2,
\end{equation}
where $Y_0$ is the particle beam half-width. Fig.\,\ref{FIG:focus}\,(a) shows typical trajectories of the electrons,
moving in the magnetic field~\eqref{eq:Buh}. The inset shows the close-up of the focus region. Initial momentum
components are equal to $p_{0,y} = 0$ and $p_{0,z} = 50$. The $y$ and $z$ coordinates at $t = 0$ are $y_0\in [-0.75, 0.75]$
and $z_0 = 0$. The oscillation frequency is $\omega_0 = \pi/10$ and the parameter $\kappa$ approximately equals
$0.157\,\omega_0^2$. The further the particles are from the axis at $t = 0$, at the larger distance along the
$z$-direction they intersect the axis. This is an example of positive spherical aberration.

In the short focusing system case, the capillary length is substantially smaller than the focus length. As a result
of the moving in the magnetic field the particle acquires the transverse velocity
\begin{equation}
  y'_0 = -\frac{Z}{c}\left(\omega_0^2 y_0 - \kappa y_0^3\right),
\end{equation}
where $Z$ is the capillary length and $y_0$ is the particle coordinate at $t = 0$. Since the particle trajectory in the
region outside of the capillary is given by $y = y_0 + y'_0 z/c$ we can find the axis intersection coordinate, which is
equal to
\begin{equation}
z_f = -c\frac{ y_0}{y'_0}\approx \frac{c^2}{Z\omega_0^2}\left(1 + \frac{\kappa}{\omega_0^2}\,y_0^2\right)
\end{equation}
i.e. the focus width is
\begin{equation}
  \Delta z \approx \frac{c^2\kappa}{Z\omega_0^4}\,Y_0^2.
\end{equation}

In Fig.\,\ref{FIG:focus}\,(b) we present the trajectories of the electrons moving in the magnetic field~\eqref{eq:Buh},
when the capillary is of a finite length equal to $Z = 2.5$. The inset shows the close-up of the focus region. Other
parameters are the same as for the case shown in Panel~(a). Here we also have an example of positive spherical aberration.

\section{Scaling\label{sec:SCALING}}

\subsection{Capillary discharge parameters\label{subsec:cap_params}}

From the theoretical model, formulated in the previous sections, the scaling of the key capillary discharge
parameters follows. The scaling provides relationships between the discharge plasma parameters: the electric
field $E$, resistance ${\cal R}$, magnetic field $B$, plasma density $n_e$, temperature $T_e$ and electric
current $I_0$, characterizing the external electric circuit, and the capillary size $L$. This yields for the
electric field
\begin{equation}\label{eq:e-pr}
  E\,[\mathrm{kV/cm}] = 0.0762\,\frac{{I_0\,[\mathrm{kA/mm}]}^{2/5}}{L\,[\mathrm{mm}]}.
\end{equation}
The resistance per unit length of the discharge, $\cal R$, depends on $I_0$ on and $L$ as
\begin{equation}\label{eq:R-pr}
  {\cal R}\,[\mathrm{Ohm}] = 0.00762\,\frac{1}{L\,[\mathrm{mm}]I_0\,[\mathrm{kA/mm}]^{3/5}}.
\end{equation}
For the electron temperature and density we have
\begin{equation}\label{eq:T-pr}
  T_e(y = 0)\,[\mathrm{eV}] = 7.58\,I_0\,[\mathrm{kA/mm}]^{2/5}
\end{equation}
and
\begin{equation}\label{eq:ne-pr}
  n_e(y = 0) = 0.8\,\langle n_e\rangle,
\end{equation}
respectively. Here the mean plasma density $\langle n_e\rangle$ is defined by Eq.\,\eqref{nemean}. The magnetic
field gradient dependence on $I_0$ on and $L$ is given by
\begin{equation}\label{eq:B_prime-pr}
  \partial_y B\bigl|_{y=0}\,[\mathrm{kGs/mm}] = 7.87\,I_0\,[\mathrm{kA/mm}]\cdot L\,[\mathrm{mm}]^{-1}.
\end{equation}

Here we considerd that, under the conditions considered in the present paper, the coefficients $\kappa_0$ and $\sigma_0$
are almost constant. They were calculated for the electron density equal to $n_e = 1.5\cdot 10^{18}\,\mathrm{cm^{-3}}$
and for the electron temperature equal to the ion temperature $T_e = T_i=3.8$\,eV. These coefficients are assumed to be
constant within the framework of the theoretical model, presented above.

The correspondence between the predictions of the theoretical model and the simulation results is due the fact that
the characteristic time for the capillary discharge to reach the equilibrium is substantially smaller than the timescale
of the electric current in the electric circuit, which is approximately equal to 100--200\,ns. Our estimates show that
characteristic time of establishing of the mechanical equilibrium $\tau_{hy}$ in $y$-direction is of the order of 6\,ns;
the time for the thermal equilibration in $y$-direction, $\tau_t$, is approximately 22\,ns; and the time for establishing
the equilibrium distribution of electric over $y$-direction $\tau_{By}$ is about 1\,ns. A characteristic time for
establishing the constant electric field over the larger axis of the capillary slot is of the order of
\begin{equation}
  \frac{\cal L}{\cal R} \approx 0.4\,\frac{L\,[\mathrm{mm}]}{R\,[\mathrm{Ohm}]}\,[\mathrm{ns}],
\end{equation}
which for the parameters under consideration is approximately equal to 35\,ns. Here $\cal L$ is inductance of the discharge
per its unit length, and $\cal R$ is the resistance of the discharge per unit of its length. A mechanical equilibrium
in $x$-direction is established in about $\tau_{hx} = 10\,\tau_{hy}\sim 60$\,ns.



In Section\,\ref{sec:SIMPLEmodel} we presented the analytical model of the plasma equilibrium inside rectangular capillaries.
The model is based on the assumption that the plasma and magnetic field parameters depend only on $y$-coordinate in
the region inside the capillary far enough from its short sides. With this model we obtained for the key parameters
the scalings given by Eqs.~\eqref{eq:e-pr}-\eqref{eq:B_prime-pr}. The results of the simulations with 2D MHD codes
of plasma and magnetic field evolution in the capillary discharge show that  the plasma reaches the equilibrium at
$t \approx 240$\,ns. The coefficients in Eqs.~\eqref{eq:e-pr}-\eqref{eq:B_prime-pr}, which were obtained from simulations,
can be compared with the ones obtained from the analytical model in Sec.\,\ref{sec:SIMPLEmodel}. The comparison is
presented in Tab.\,\ref{T1}.
\begin{table}[tbph]
\caption{Comparison of the coefficients in the scaling given by
Eqs.~\protect\eqref{eq:e-pr}-\protect\eqref{eq:B_prime-pr} with those obtaned in the MHD simulation.}
\begin{tabular}{c|cc}
\toprule
& ~~~simulation~~~ & ~~~theory~~~ \\
\cline{1-3}
E
& 0.0694 &0.0762\\
${\cal R}$
& 0.00694 &0.00762\\
$T_e$
& 7.10 &7.58\\
$n_e$
& 0.867 &0.798\\
$\partial_y B$
& 7.28 &7.87\\
\botrule
\end{tabular}
\label{T1}
\end{table}

The relative accuracy of the analytical model prediction of the quasi-stationary stage parameters compared to the
simulation results is of the order of 10\%. Discrepancies of the same order take place in the coefficients describing
spatial distribution of $n_e$ and $B_x$. They can be derived by comparing of Eqs.~\eqref{3_010} and~\eqref{3_020} with
Eqs.~\eqref{eq:nexi3} and~\eqref{eq:Buh}, respectively. These minor discrepancies are caused by several reasons.
The simulations take into account the weak non-stationarity of the discharge even in the quasi-stationary stage,
the inhomogeneous distribution of the electric current, as well as weak plasma motion. Another source of minor
inaccuracy of the analytic model of the same order is replacement of the Coulomb logarithms by constant values.

\subsection{Focusing of charged particle beams}

In the homogeneous magnetic field the charged particle rotates along the circle of the Larmor radius equal to
$r_B = pc/eB$, where $p$ is the particle momentum. If the magnetic field is inhomogeneous and is vanishing at
the surface $y = 0$ as in the case considered above, $B = h_1 y/L$, the charged particle trajectory does not
intersect the $y = 0$ plane, being localized in the region approximately equal to
$r_B(\bar y) = pc/eB(\bar y) = pcL/eh_1\bar y$ provided the mean trajectory distance from the null plane $y = 0$
is substantially larger than $r_B(\bar y)$. The magnetic field inhomogeneity results in the charged particle motion
along the $z$-axis due to the gradient drift. In the limit $\bar y\to 0$ the particle Larmor radius formally tends to
infinity, nontheless the trajectory is localized in the finite size region in the vicinity of the neutral plane $y = 0$.
One can find the space scale of the particle trajectory localization in the vicinity of the $y = 0$ plane, $y^*$,
using the relationship $y^* = r_B(y^*)$:
\begin{equation}
y^* = \sqrt{pcL/eh_1} = 0.9\sqrt{pcL/eB_m} \,.
\end{equation}
Here we take into account, that  $h_1$ and maximum magnetic field $B_m$ given by Eq. \eqref{eq:Bmax} are related as 
$  h_1=B_m /(\sqrt{{7}/{5}}\, u_0^{2/7})$
with $u_0=0.25$.
The charged particle with the initial coordinate $y_0$ larger than $y^*$” undergoes the so-called gradient drift 
directed perpendicularly to the magnetic field 
and to the magnetic field gradient, i. e. directed along the $z$ axis. The beam in which the particle initial 
coordinates $y_0$ are outside of the interval 
$[- y^*, y^*]$ cannot be focused. Instead, if 
 the charged particle initial coordinate $y_0$ is significantly less than $y^*$ the trajectory is localized within
the region along the $y$ axis of the size approximately equal to $y_0$ intersecting the $y = 0$ plane at the distance
$z_f \approx y^*$. We notice that the initial coordinates of the trajectories presented in Fig. 7 are chosen to be 
inside the interval $[- y^*, y^*]$.

The scaling of the charged particle focusing characterized by the dependence the focusing distance $z_f$ on the magnetic
field parameters and charged particle energy can be considered in two limits.

In the ultra relativistic limit, assuming that the charged particle is an electron of the energy $m_e c^2 \gamma_e$,
we find
\begin{equation}
  z_f = 0.9\,\sqrt{\frac{L\,m_e c^2 \gamma_e}{e B_m}}.
\end{equation}
For example, for $B_m = 10$\,kGs and $L = 1$\,mm, that corresponds to current line density of 16\,kA/cm, and
 $\gamma_e = 2$ (i.e. for the electron with the energy of 1\,MeV) it is approximately equal
to 0.17\,cm. If the electron energy is of 10\,GeV (i.e. $\gamma_e = 2\cdot 10^4$) and $B_m = 10$\,kGs and $L = 1$\,mm
the focusing distance is equal to 17\,cm.

In the case of the non-relativistic limit, assuming that it is an ion of the mass $m_i = A_i m_p$, where $A_i$ is
the ion atomic number and $m_p$ is the proton mass, and of the electric charge $Z_i e$, the focusing distance can
be written as
\begin{equation}
  z_f = 0.9\,\sqrt{\frac{L\, A_i m_p c^2 \beta_i}{Z_i e B_m}}\,,
\end{equation}
where $\beta_i = v_i/c = \sqrt{1 - 1/({\cal E}_i/m_i c^2 + 1)^2}\approx \sqrt{2{\cal E}_i/m_i c^2}$ is a normalized
velocity of the ion with the energy ${\cal E}_i$. For the 100\,MeV proton ($A_i = 1$ and $Z_i = 1$) moving inside the
$L = 1$\,mm size capillary with the 1\,kGs magnetic field, corresponding to current line density of 1.6\,kA/cm, the
focusing distance is equal to 10.8\,cm. If the ion is carbon with atomic number $A_i = 12$, charge $Z_i = 6$, and energy
${\cal E}_i = 1$\,GeV moving in the $L = 1$\,mm capillary magnetic field $B_m = 10$\,kGs the focus is at the distance
$z_f = 4.6$\,cm.

As we see, with the capillary plasma generated magnetic field the focusing length is substantially less than that in
the case of the focusing systems used in the standard charged particle accelerator technology both for multi GeV
electrons and multi hundred MeV ions.

\section{Conclusions\label{sec:CONCLUSIONS}}

We suggest a novel method for the asymmetric focusing of electron beams. The scheme is based on the active lensing
technique, which takes advantage of the strong inhomogeneous magnetic field generated inside the capillary discharge
plasma. The plasma and magnetic field parameters inside a capillary discharge are described theoretically and modeled
with dissipative magneto-hydrodynamic simulations to enable analysis of capillaries of oblong rectangle cross-sections
implying that large aspect ratio rectangular capillaries can be used to form flat electron and/or ion bunches. In the
case of flat electron bunches the beamstrahlung in linear colliders or storage rings weakens. The effect of the
capillary cross-section on the charged particle beam focusing properties is studied using the analytical methods
and simulation-derived magnetic field map showing the range of the capillary discharge parameters required for
producing the high quality flat electron beams. Active lensing of charged particle beams with the inhomogeneous
magnetic field generated in the capillary discharge plasma will enable developing of compact laser plasma based
systems for accelerating high energy charged particles and for manipulating the beam parameters.

\section*{ACKNOWLEDGMENTS}

The work was supported in part by the Russian Foundation for Basic Research (Grant No.\,15-01-06195),
by the Competitiveness Program of National Research Nuclear University MEPhI (Moscow Engineering Physics Institute),
contract with the Ministry of Education and Science of the Russian Federation No.\,02.A03.21.0005, 27.08.2013
and the basic research program of Russian Ac. Sci. Mathematical Branch, project 3-OMN RAS.
The work at Lawrence Berkeley National Laboratory was supported by US DOE under contract No.\,DE-AC02-05CH11231.
The work at KPSI-QST was funded by ImPACT Program of Council for Science, Technology and Innovation (Cabinet Oce,
Government of Japan).
At ELI-BL it has been supported by the project ELI~-- Extreme Light
Infrastructure~-- phase\,2 (CZ.02.1.01/0.0/0.0/15 008/0000162) from European Regional Development Fund,
and by the Ministry of Education, Youth and Sports of the Czech Republic (project No.\,LQ1606) and by the
project High Field Initiative (CZ.02.1.01/0.0/0.0/15\_003/0000449) from European Regional Development Fund.

\end{document}